\documentclass[preprint,aps,showpacs]{revtex4}

\newcommand{\omits}[1]{}

\begin{document}

\preprint{hep-th/0310141}

\title{
(Super)gravity and Yang-Mills Theories\\
as Generalized Topological Fields with Constraints}

\author{Yi Ling${}^{1,2}$}
\email{yling@itp.ac.cn}
\author{Roh-Suan Tung${}^{2}$}
\email{tung@itp.ac.cn}
\author{Han-Ying Guo${}^{1,2}$}
\email{hyguo@itp.ac.cn}
\affiliation{%
${}^1$ CCAST (World Laboratory), P.O. Box 8730, Beijing
   100080, China,}

\affiliation{%
${}^2$ Institute of Theoretical Physics,
 Chinese Academy of Sciences,
 P.O.Box 2735, Beijing 100080, China\footnote{Mailing address.}.}

\date{\today}

\begin{abstract}
\quad We present a general approach to construct a class of
generalized topological field theories with constraints by means
of generalized differential calculus and its application to
connection theory. It turns out that not only the ordinary $BF$
formulations of general relativity and Yang-Mills theories, but
also the $N=1,2$ chiral supergravities can be reformulated as
these constrained generalized topological field theories once the
free parameters in the Lagrangian are specially chosen. We also
show that the Chern-Simons action on the boundary may naturally be
induced from the generalized topological action  in the bulk,
rather than introduced by hand.
\end{abstract}



\maketitle

\newcommand{\GForm}[2]{\buildrel\scriptstyle {#1} \over {\bf #2}}
% Example: $\GForm{p}{a}$
%%%%%%%%%%%%%%%%%%%%%%%%%%%%%
% Regular p-forms
\newcommand{\Form}[2]{\buildrel\scriptstyle {#1} \over #2}
% Example: $\Form{p}{\alpha}$
%%%%%%%%%%%%%%%%%%%%%%%%%%%%
% GDC-d
\newcommand{\Gd}{{\bf d}}
% Example: \d
%%%%%%%%%%%%%%%%%%%%%%%%%%%%
%\tableofcontents
%%%%%%%%%%%%%%%%%%%%%%%%%%%%
\newpage
\section{Introduction}

 Recent progress in the
study of quantum theory of gravity
in terms of Ashtekar-Sen variables \cite{AS} has %es %in non-perturbative quantum gravity
 revealed that incorporating the methods of topological quantum
field theory into non-perturbative quantum gravity may push this
approach forward significantly (for a recent review, see, e.g.
\cite{Smolin2003}). The starting point is that the classical
Lagrangian of gravity theories in the first order formalism can
always be written as constrained topological field theories. Such
an extraordinary example is the Plebanski-like Lagrangian for
general relativity (GR) with cosmological constant
\cite{Plebanski}. It has the form of $BF$ theory with appropriate
gauge group  in four dimensions
\begin{equation}\label{BF1}
S_{BF}(A,B)=\int_{ M}Tr(B\wedge F+{\Lambda\over 12}B\wedge
B).\end{equation} As a matter of fact, such a framework can be
extended to construct (at least $N\le 2$) supergravities in fours
dimensions \cite{super-ezawa}, as well as $D=11$ supergravity
\cite{LS}.

Very recently, we have shown \cite{BFT} that based on the
Generalized Differential Calculus ($GDC$)
\cite{Robinson,NR2001}(see also the appendix), a generalized
connection theory may be established. Furthermore, Einstein's
general relativity can be reformulated from a kind of generalized
topological field theory ($GTFT$).  The $GDC$ originally was
advocated to study the twistor theory.  In this context
 an ordinary $p$-form on a manifold $M$ is extended to a
generalized $p$-form defined by an ordered pair of the original
$p$-form and an additional $(p+1)$-form on the same manifold.
Correspondingly,
 the ordinary exterior differential $d$ is replaced by the
generalized exterior differential $\Gd$ that is also nilpotent.
 Thus when
applying this $GDC$ to the principle bundle $P(M,G)$ on $M$ with
gauge group $G$, the ordinary $g$-valued connection 1-form $A$ and
curvature 2-form $F$ are generalized to a $g$-valued generalized
connection 1-form $\bf A$ and a curvature 2-form $\bf F$,
respectively. By definition, $\bf A$ is a $g$-valued pairing of
$(A, B)$ with $B$ an ordinary $g$-valued gauge covariant 2-form,
while $\bf F$ is the pairing $(F+kB, D_A B) $ with $k$ an
arbitrary constant and $D_A$ the ordinary covariant derivative
with respect to $A$. It has also been shown in \cite{BFT} that
with respect to $\Gd$, $\bf F$ satisfies the Bianchi identity and
its gauge invariant polynomials  are endowed with the Chern-Weil
homomorphism. Since there is an arbitrary gauge covariant 2-form
$B$ in the definition of $\bf A$,  this framework may combine an
ordinary topological term in topological field theory ($TFT$) with
other terms as the parts in $GTFT$ with particular chosen field
$B$. In addition, it is interesting to see that this kind of
$GTFT$ also leads to certain relation between the ordinary local
Chern-Simoms term on the boundary $\partial M$ and the
BF-Lagrangian in the bulk $M$ as was shown in \cite{BFT}.

In this paper, we intend to develop a general approach to
reformulate other gravity theories, particularly the $N=1,2$
chiral supergravities, as well as pure Yang-Mills theory together
with corresponding topological term, respectively, as this sort of
the constrained $GTFTs$. Namely, we  show that the ordinary
Lagrangian of gravity theories together with
 ordinary topological terms
may be simply written as a ${\it generalized}$ topological term
with appropriate constraints. This generalized topological action
has a
 form of the integral of the generalized second Chern class as
 Lagrangian on four dimensions
\begin{equation} S=\int_{{ M}}STr({\bf F}\wedge {\bf
F}),\end{equation} where $STr$ is understood as the supertrace in
the case of supergravities. This action includes both the ordinary
(topological) second Chern class  $STr(F\wedge F)$ and $BF$-type
terms as well as their super-partners in the case of
supergravities. Thus, this offers a framework of $GTFT$ with
constraints for specially chosen $B$ or $B^s$ in the case of
gravity and Yang-Mills or supergravities, respectively. It is
worthwhile to mention that  this formalism  provides a very
convenient framework for constructing the ($N=1,2$)
supergravities, since it combines all components of the theory
into a compact manner so as to the gauged Lie super-algebra is
automatically closed.

The paper is arranged as follows. In section two we present the
general formalism of $GTFT$ based on the generalized
(super)-connection theory. Its  topological features may be
formally understood by considering the generalized Chern-Weil
homomorphism for generalized curvature invariant polynomials. Then
we exhibit the geometric formulation of Einstein-Hilbert action
and the Yang-Mills theory on curved spacetime as this kind of
$GTFTs$ with constraints in section three. In section four and
five, we turn to show that $N=1$ and $N=2$ chiral supergravities
with some free parameters can also be formulated as the $GTFT$
with constraints, respectively. In all these cases, we find that
the BF-type action in the bulk is always as a counterpart of the
ordinary topological terms, which may naturally induce
Chern-Simons-type action on the boundary $\partial M$. We end this
paper by some remarks focusing on its possible applications to the
quantization of gravitational field, which should be further
considered in the future.

\section{General Formalism for the Generalized Topological Field Theory}

In this section, we first recall the properties of the generalized
gauge fields \cite{BFT} in the framework of $GDC$
\cite{BFT,Robinson,NR2001}. Then we  %will
generalize it to the supersymmetric case. Namely, we introduce the
Lie super-algebra $g^s$ valued gauge fields as well as their gauge
invariant generalized curvature polynomials.

We adopt the conventions and notations through this paper as
follows. The Lie super-algebra $g^s$-valued connections and
curvatures are written as script letters, while the generalized
super-connection and super-curvature in the context of $GDC$ are
written as the bold letters. The upper-case Latin letters
$A,B,...=0,1$ denote two component spinor indices, which are
raised and lowered with the constant symplectic spinors
$\epsilon_{AB}=-\epsilon_{BA}$ together with its inverse and their
conjugates according to the conventions
$\epsilon_{01}=\epsilon^{01}=+1$,
$\lambda^A:=\epsilon^{AB}\lambda_B$, $\mu_B:=\mu^A\epsilon_{AB}$
\cite{PR}.

Let us consider a principle bundle $P(M^4, G)$ over the
4-dimensional manifold $M^4$ with a semisimple gauge group $G$ as
the structure group. A Lie algebra $g$-valued generalized
connection ${\bf A}$ is defined as a $g$-valued pairing of a
$g$-valued ordinary connection 1-form $A$ and a $g$-valued
ordinary 2-form $B$
\begin{equation}
{\bf A}=( A^{p}, B^{p} ) T_{p} =(A, B),\quad T_p \in g, \label{1}
\end{equation}
where $B$ is assumed as gauge covariant under the gauge
transformations in order to introduce a $g$-valued generalized
curvature 2-form ${\bf F}$ that is gauge covariant
\begin{eqnarray}
 {\bf F} &=&\Gd{\bf A} + {\bf A} \wedge {\bf A}\nonumber\\
 &=& \left( dA+A\wedge
 A+k B, \quad  d B+
 A\wedge B -B \wedge A \right) \nonumber\\
 &=&(F+ k B,~ D B),
\end{eqnarray}
where $D=D_A$ is the covariant derivative with respect to the
connect $A$, $k$ is an arbitrary constant. It is easy to show that
the $g$-valued generalized curvature 2-form ${\bf F}$ satisfies
the Bianchi identity via $GDC$:
\begin{eqnarray}\label{GBI}
 {\bf D} {\bf F} &:=& \Gd {\bf F}
 + {\bf A}\wedge
 {\bf F} - {\bf F} \wedge {\bf A}
 \nonumber \\
 &=&( DF, ~D^2 B) \equiv 0 .
\end{eqnarray}

 In \cite{BFT}, we establish the generalized Chern-Weil homomorphism
for generalized curvature invariant polynomials via $GDC$ on any
even dimensional manifolds. But their topological meaning should
be as same as the original curvature invariant polynomials. The
reason is that the cohomology with respect to $\Gd$ in $GDC$ is
trivial, which is easy to be proved (see the appendix).

Now we may introduce an action on the base manifold $M^4$ as the
topological form with the generalized second Chern-class as
Lagrangian for the generalized curvature 2-form:
\begin{eqnarray}\label{a}
{\cal S}_T=\int_{M^4}{\cal L}_T=\int_{M^4} Tr ({\bf F}\wedge {\bf
F}) = \int_{M^4} \Gd{\bf Q}_{CS}.
\end{eqnarray}
Here ${\bf Q}_{CS}$ is the generalized local Chern-Simons 3-form,
i.e., the  pairing of a 3-form and a 4-form
\begin{eqnarray}\label{CS}{\bf Q}_{CS}
&=& Tr ({\bf A} \wedge {\bf F}-\textstyle{1\over3}
{\bf A} \wedge {\bf A} \wedge {\bf A})\nonumber\\
&=&  Tr (A \wedge F -\textstyle{1\over3} A \wedge A \wedge A
+k A\wedge B, \nonumber\\
&&\qquad ~ A \wedge DB+B\wedge F+k B\wedge B ).
\end{eqnarray}
Note that in the pairing of the local generalized Chern-Simons
term above, the 3-form is the usual Chern-Simons term up to an
$k\, Tr(A\wedge B)$ term while the 4-form is the usual BF term up
to a $Tr (A\wedge DB)$ term. Namely, although the topological
meaning of the generalized second Chern class is the same as
before, the generalized local Chern-Simons term combines the
ordinary Chern-Simons 3-form and the BF 4-form together as an
entire object via $GDC$. This already leads to a relation between
the Chern-Simons term and the BF-term.

In fact, the generalized second Chern class as a Lagrangian 4-form
in (\ref{a}) is a pairing of a 4-form and a 5-form:
\begin{eqnarray*}
{\cal L}_{\rm{T}}%[{\cal A}^p]
&=& Tr ( F\wedge F + 2k B\wedge F+k^2 B \wedge B,
\nonumber\\
&&\qquad \quad 2(F \wedge DB+k B\wedge DB) ).
\end{eqnarray*}
Using the Bianchi identity, we can rearrange the 5-form so that
\begin{eqnarray}\label{a1}
{\cal L}_{\rm{T}}&=& Tr ( F \wedge F + 2k B\wedge F+k^2
B\wedge B, \nonumber\\
&&\qquad ~ d \left(2 B \wedge F+k B\wedge B\right) ).
\end{eqnarray}
The first term is just the ordinary topological term, i.e. the
second Chern class $Tr(F\wedge F)$, together with BF Lagrangian as
its counterpart appeared via $GDC$. And the second term is a total
derivative of the BF Lagrangian.

Once more, the pairing of the action (\ref{a}) shows a relation
between two types of $TFTs$, the Chern-Simons-type on the boundary
and the BF-type in the bulk on four dimensions:
\begin{eqnarray}
{\cal S}_{\rm{T}}[{\bf A}]&=&\int_{M^4} {\cal L}_T=\int_{M^4} \Gd{\bf Q}_{CS} \nonumber\\
 &=& \int_{M^4}  Tr \left( F\wedge F + 2 k B \wedge F + k^2 B\wedge
 B \right).
 \end{eqnarray}
In addition, although as a five form, the total derivative of the
BF Lagrangian in (\ref{a1}) vanishes automatically on the
4-dimensional manifold $M^4$, it also indicates that the BF
Lagrangian on four dimensions may be regarded as a term on the
boundary of a topological action (\ref{a}) on a five dimensional
buck. This implies that there may exist a kind of  decent
relations among these topological terms on different dimensions.
We will leave this topic for forthcoming publications.

Let us now consider the supersymmetric generalization of the above
issues. In fact, if we still consider the principle bundle $P(M^4,
G^s)$ with a Lie super-group $G^s$ as the structure group, it is
more or less straightforward to generalize the above issues to the
supersymmetric case as long as the Lie algebra $g$ is replaced by
a Lie super-algebra $ g^s$ of $G^s$.

For the sake of  explicitness, we may take the Lie super-algebra
$Osp(2|N)$ as an example. The ordinary super connection 1-form
valued on this Lie super-algebra is defined as,
\begin{equation}
{\cal A}=A^{AB}J_{AB}+\psi^{I A}Q_{I A}+A^{IJ}Z_{IJ},
\end{equation}
where $J_{AB}, Q_{I A} $ are the bosonic, fermionic generators of
the algebra, respectively, and $Z_{IJ}$ %are
the generators of automorphism group $SO(N)$. The indices $IJ$ are
antisymmetric and may run from 1 to N, depending on the order of
supersymmetry.

In order to apply the GDC to the case of super-algebra, we
introduce an ordinary 2-form fields valued on $g^s$, which should
be gauge covariant,
\begin{equation}
{\cal B}=B^{AB}J_{AB}+B^{IA}Q_{IA}+B^{IJ}Z_{IJ},
\end{equation}
and define a generalized connection 1-form as
\begin{equation}
{\bf A^s}=({\cal A}, {\cal B})=(A^{AB}, B^{AB})J_{AB}+(\psi^{IA},
B^{IA})Q_{I A}+(A^{IJ}, B^{IJ})Z_{IJ},
\end{equation}
where the upper index $s$ denotes that the object is a Lie
super-algebra valued. In what follows, it is often omitted for the
sake of simplicity.

Following the rules of GDC, we find that the Lie super-algebra
valued generalized  curvature 2-form may be defined as
\begin{eqnarray}
{\cal \bf F} &=& {\bf d}{\cal\bf A}+{\cal \bf A}\wedge {\cal\bf A}\nonumber\\
         &=& (F^{AB}+kB^{AB}-{a\over 2}\psi^{IA}\wedge \psi_I^B, DB^{AB}-a\psi^{IA}\wedge B_I^B)J_{AB}\nonumber\\
         && + (F^{I A}+kB^{I A}+{a\over 2}A^{IJ}\wedge \psi_J^A, DB^{I A}
         +a A^{IJ}\wedge B_J^A)Q_{I A}\nonumber\\ &&+(F^{IJ}+kB^{IJ}+{1\over 2}\psi^{IA}\wedge\psi^J_A,DB^{IJ}+\psi^{IA}\wedge B^J_A)Z_{IJ},
\end{eqnarray}
where
\begin{eqnarray}
F^{AB}&=& dA^{AB}+A^{AC}\wedge {A_C}^{B},\;\;\;\;\;\;\; F^{I
A}=d\psi^{I A}+A^{AB}\wedge \psi^I_B,\nonumber\\ F^{IJ}&=&
dA^{IJ}+A^{IK}\wedge A_K^{J},\end{eqnarray} and $a$ is some
coupling constant appearing in the superalgebra. It is
straightforward to show that this Lie super-algebra valued
generalized curvature satisfies the Bianchi identity:
\begin{eqnarray}
 {\cal \bf D} {\cal \bf F} &:=& \Gd {\cal\bf F}
 + {\cal\bf A}\wedge
 {\cal\bf F} - {\cal \bf F} \wedge {\cal\bf A}
 \nonumber \\
 &=&( DF^{AB}, D^2 B^{AB})J_{AB}+(DF^{I A},D^2B^{I A})Q_{I A}+(\tilde{D}F^{IJ},\tilde{D}^2B^{IJ})Z_{IJ} = 0 .
\end{eqnarray}

Now it is also easy to show that the formal topological meaning
for the gauge invariant polynomial, ${\cal P}({\cal\bf F})$, of
the Lie super-algebra valued generalized curvature 2-forms.
Namely, it is closed with respect to $\Gd$ and satisfies the
Chern-Weil homomorphism:
\begin{eqnarray}
(i) \qquad&& \Gd{\cal P}({\cal\bf F})=0 , \\
(ii) \qquad&& {\cal P}({\cal\bf F}_1)-{\cal P}({\cal\bf
F}_0)=\Gd{\cal Q}({\cal\bf A}_0, {\cal\bf A}_1), \label{cs1}
\end{eqnarray}
where ${\cal Q}({\cal\bf A}_0, {\cal\bf A}_1)$ is the secondary
Chern-Simons invariant polynomial for the generalized curvature.

Now, we are ready to introduce the action as the integral of the
generalized second Chern class:
\begin{equation}\label{SFF-1}
S=\int_{ M}STr{\cal\bf F}\wedge{\cal\bf F}.\end{equation}

The formal topological character of the action (\ref{SFF-1}) can
be understood by investigating the above Chern-Weil homomorphism
formula as we did in \cite{BFT}. In particular, the Lagrangian
4-form ${\cal L}$ can be given by taking ${\cal P}({\cal\bf F})$
as the second Chern class for the  Lie super-algebra valued
generalized curvature, as well as ${\cal\bf A}_1={\cal\bf A}$ and
${\cal\bf A}_0=0$ in (\ref{cs1}), then
\begin{eqnarray}\label{GCW}
STr ({\cal\bf F}\wedge {\cal\bf F})=\Gd{\cal Q}_{CS},\label{QCS}
\end{eqnarray}
where ${\cal Q}_{CS}$ is the generalized local Chern-Simons
3-form, i.e., the  pairing of a 3-form and a 4-form
\begin{eqnarray}\label{CS2}{\cal Q}_{CS}
&=& STr ({\cal\bf A} \wedge {\bf d} {\cal\bf
A}+{\textstyle{2\over3}}
{\cal\bf A} \wedge {\cal\bf A} \wedge {\cal\bf A})\nonumber\\
&=& ( STr ({\cal A} \wedge d{\cal A} + {\textstyle{2\over3}} {\cal
A} \wedge {\cal A} \wedge {\cal A}
+k {\cal A}\wedge {\cal B}),  \nonumber\\
&& \qquad ~ STr( {\cal A} \wedge D{\cal B}+{\cal B}\wedge {\cal
F}+k {\cal B}\wedge {\cal B}) ).
\end{eqnarray}
Furthermore given the specific bilinear form of the supertrace for
a superalgebra the super Chern-Simons terms can be written in
components.

Thus, we have established the general formalism for a kind of
$GTFTs$ and their supersymmetric generalizations via GDC.
Significantly, it is straightforward to see that in the
supersymmetry cases of the $GTFT$ the ordinary super-topological
term in the action is also companied by the super-BF-type term and
there is a relation between the super-BF-type term in the bulk of
$M$ and the super-Chern-Simons term on the boundary $\partial M$.

\section{Einstein-Hilbert Action and Yang-Mills Action as Constrained BF Gauge Theories}

In this section, we first briefly recall how to reformulate the
Einstein-Hilbert action together with corresponding topological
term as constrained BF gauge theories in the framework of $GTFT$.
Then we show that the  Yang-Mills theory on the curved spacetime
can also be reformulated as a $GTFT$ with constraints.

\subsection{Einstein-Hilbert Action}

For GR, it is well know that the gauge group is the homogeneous
Lorentz group or its covering group $SL(2,C)$. Consider the
$sl(2,C)$ algebra:
\begin{eqnarray}
\left[ J_{AB}, J_{CD} \right] &=&
      \epsilon_{C(A} J_{B)D}
     +\epsilon_{D(A} J_{B)C} ,
\end{eqnarray}
where $\epsilon_{C(A} J_{B)D} =\textstyle{1\over2} (\epsilon_{CA}
J_{BD} +\epsilon_{CB} J_{AD})$. The local Minkowskian metric on
the tangent space $\eta_{pq}={\rm{diag}}(\eta_{(AB)(MN)})$ is
given by
\begin{eqnarray}
\eta_{(AB)(MN)}&=& \textstyle{1\over2}
(\epsilon_{AM}\epsilon_{BN}+\epsilon_{AN}\epsilon_{BM}).
\end{eqnarray}

We introduce an $sl(2,C)$-valued generalized connection 1-form on
the tangent bundle $P(M^4, SL(2,C))$ via GDC,
\begin{equation}
{\bf A}=(A^{AB}, B^{AB} ) J_{AB} , \label{71}
\end{equation}
where $A^{AB}$ is the ordinary $sl(2,C)$-valued connection 1-form
on the bundle and $B^{AB}$ is an $SL(2,C)$-gauge covariant 2-form.

Given such a connection ${\bf A}$, the generalized curvature
2-form ${\bf F}={\bf F}^p T_p={\bf
F}^{AB} J_{AB}$ can be given via $GDC$ with the components% by
\begin{equation}
{\bf F}^{AB}=(F^{AB}+ k B^{AB},~~ D B^{AB}) .
\end{equation}
It does satisfies the Bianchi identity (\ref{GBI}) via $GDC$.

A simple generalized topological Lagrangian  4-form in (\ref{a})
of this curvature $\bf F$ is the second Chern class-like class as
follows:
\begin{eqnarray}
{\cal S}_{\rm{SL(2,C)}}[{\bf A}]&=&\int_{M^4} {\cal
L}_{\rm{SL(2,C)}}
=\int_{M^4} \, {\bf F}^{AB}\wedge {\bf F}_{AB} +c.c. \nonumber\\
&=& \int_{M^4} \,( R^{AB} \wedge R_{AB} + 2k R^{AB}\wedge
B_{AB}+k^2 B^{AB} \wedge B_{AB}) +c.c. \label{SL2C-action}
\end{eqnarray}

It should be noted that first the second term in the above action
is almost the same as the BF-type action in (\ref{BF1}). Secondly,
the 4-form Bianchi identity (the second term) in (\ref{GBI}), $D^2
B^{AB}=0$, looks identical to the Bianchi identity,
$D^2(e^{AA'}\wedge e^{B}{}_{A'})=0$, of the first Cartan structure
equation for ordinary torsion 2-form, $T^{AA'}:=De^{AA'}$. Thus we
may introduce 1-form fields $e^{AA'}$ to parameterize the tangent
space of $M$, which is the coset $Sp(4)/SL(2,C)$,
and take $B^{AB}$ to be %equal to
$ l^{-2} e^{AA'}\wedge e^B{}_{A'}$ with $l$ a dimensional constant
\footnote{ We will ignore this dimensional constant as well as the
Newton constant through this paper for convenience. But since
$e^{AA'}$ is dimensionless while $\psi^A$ has the dimension
$l^{-1/2}$, it's easy to recover all these dimensional factors in
the Lagrangian. }.

The $sl(2,C)$-valued connection generalized 1-form now becomes
{\footnote{The formulation can be written with purely unprimed
spinors by defining spinor 1-forms $\varphi^A=e^{A0'}$ and
$\chi^A=e^{A1'}$ \cite{TJ}. In terms of these spinor 1-forms, the
purely unprimed $sl(2,C)$-valued generalized connection 1-form
 is ${\cal A}^+=\left(\omega^{AB}, ~~(2 / l^2)~
\chi^{(A} \wedge \varphi^{B)} \right) M_{AB}$ .}}
\begin{equation}
{\bf A}=(A^{AB},  e^{AA'}\wedge e^B{}_{A'} ) M_{AB} +c.c.
\label{61}
\end{equation}
The action (\ref{SL2C-action}) becomes
\begin{eqnarray}
&& {\cal S}[A^{AB},A^{A'B'},e^{AA'}]=\int_{M^4} Tr({\bf F}
\wedge {\bf F})+c.c. \nonumber\\
&=& \int_{M^4}( R^{AB} \wedge R_{AB} + {2k} R^{AB}\wedge
e^{AA'}\wedge e^B{}_{A'}+ {k^2} e^{AA'}\wedge e^B{}_{A'} \wedge
e_{A}{}^{C'}\wedge e_{BC'} +c.c.),
\end{eqnarray}
where the generalized curvature is given by
\begin{eqnarray}
{\bf F}%={\cal F}^p T_p
&=&{\bf F}^{AB} M_{AB} +c.c.\nonumber\\
%\end{equation}
%where
%\begin{equation}
 {\bf F}^{AB}=(F^{AB} &+& {k} e^{AA'}\wedge e^B{}_{A'},
 D (e^{AA'}\wedge e^B{}_{A'}) )
 .
\end{eqnarray}

Thus the action (\ref{SL2C-action}) is the Einstein-Hilbert action
with  the cosmological constant and a topological term. Namely, GR
in the absence of matter may be formulated as a GTFT via GDC.

By varying with respect to $A^{AB}$, we obtain
\begin{equation}
D(e^{AA'}\wedge e^{B}{}_{A'})=0 ,\label{FE1}
\end{equation}
which gives the equation for  torsion-free. While by varying with
respect to $e^{AA'}$, we obtain
\begin{equation}
F^{AB}\wedge e_B{}^{A'}+{k} e^{AB'}\wedge e^{B}{}_{B'} \wedge
e_B{}^{A'}+c.c.=0 ,\label{FE2}
\end{equation}
which is the Einstein equation with a cosmological constant
$\Lambda:={k/ l^2}$.

Alternatively,  we can consider adding a constraint on $B^{AB}$ as
in \cite{Smolinholo,CDJM}
\begin{eqnarray}
{\cal S}[{\bf A}^p,\lambda_{AB},e^{AA'}] &=&\int_{M^4}\, {\bf
F}^{AB}\wedge {\bf F}_{AB} +\lambda_{AB} \wedge ( e^{AA'}
\wedge e^{B}{}_{A'}-B^{AB}) + c.c.\nonumber\\
&=& \int_{M^4}\, F^{AB} \wedge F_{AB} + 2k F^{AB}\wedge B_{AB}+k^2
B^{AB}\wedge B_{AB}  \nonumber\\
 &&+ \lambda_{AB} \wedge
(  e^{AA'}\wedge e^{B}{}_{A'}-B^{AB}) + c.c.,
\end{eqnarray}
where $\lambda_{AB}$ is the Lagrangian multiplier. The variational
principle leads to the same equation with (\ref{FE2}).

\subsection{Yang-Mills Action on Curved Spacetime}

Let us now consider Yang-Mills fields on the curved spacetime
$M^4$. As is well known, the BF theory provides a strategy to
write down Yang-Mills theory in a pure geometric formalism. Such a
formulation has been studied at many places, for instance
see\cite{CDJ,BFYM}. Quantizing this first-order formalism is
supposed to give more insight into the non-local and
non-perturbative features of $4D$ Yang-Mills theory. In this
section we show that this formalism can be derived from the $GTFT$
via $GDC$ as well.

 Consider a semisimple gauge group $G$ with generators
$T_p$. We introduce a $g$-valued generalized connection 1-form as
follows
\begin{equation}
{\bf A}=(A^{p}, B^{p} ) T_{p} , \label{ym1}
\end{equation}
where $A^{p}$ is the ordinary $g$-valued connection 1-form on the
bundle $P(M^4, G)$ and $B^{p}$ is a $g$-valued gauge covariant
2-form. Given such a generalized connection ${\bf A}$, the
generalized curvature ${\bf F}={\bf F}^p T_p$   is given via $GDC$
with components
\begin{equation}
{\bf F}^{p}=(F^{p}+ k B^{p},~~ D B^{p}) .
\end{equation}

A simple generalized Lagrangian  4-form using this connection $\bf
A$ is
\begin{eqnarray}
{\cal S}[{\bf A}]&=&\int_{M^4} \, {\bf F}^{p}\wedge {\bf F}_{p} \nonumber\\
&=& \int_{M^4} (\, F^{p} \wedge F_{p} + 2k F^{p}\wedge B_{p}+k^2
B^{p} \wedge B_{p}). \label{YM-action}
\end{eqnarray}
It is of the BF-type term with an ordinary second Chern-class.

Consider the constraint:
\begin{equation}
B^p=\phi^p_{AB} B^{AB},
\end{equation}
where $B^{AB}=e^{AC'}\wedge e^{B}{}_{C'}$. The action
(\ref{YM-action}) becomes
\begin{eqnarray}
&& {\cal S}_{\rm{YM}}[A,\phi_{AB},B^{AB}]=\int_{M^4} Tr({\bf F}
\wedge {\bf F})\nonumber\\
&=& \int_{M^4}\, Tr[ F \wedge F +  2k F \wedge \phi_{AB} B^{AB} +
k^2 \phi_{AB} \phi_{CD} B^{AB}\wedge B^{CD}].\label{CYM}
\end{eqnarray}
To see this action is actually the ordinary Yang-Mills theory in
curved spacetime, we may vary this action with respect to
$\phi_{AB}$, then obtain the following equation of motion
\begin{equation}
k F \wedge B^{AB} + k^2 \phi_{CD} B^{AB}\wedge B^{CD} =0,
\end{equation}
which can be solved for $\phi_{CD}$. Note that we can expand $F$
to the basis of 2-forms ($B^{CD},B^{C'D'}$)\cite{CDJ},
\begin{equation}
F=F_{CD} B^{CD} +F_{C'D'} B^{C'D'}.
\end{equation}
By comparing with the field equation for $\phi$, we obtain
\begin{equation}
\phi_{AB}=-{1\over k} F_{AB}.
\end{equation}
 It is easy to see that $-k\phi_{AB}$ is the
self-dual part of the Yang-Mills curvature 2-form $F^p$ and
\begin{equation}
B^p=\phi^p_{AB} B^{AB}= -{1\over 2 k}(1-*) F^p .
\end{equation}
Inserting it into the action then yields
\begin{equation}
S_{YM}=\int_{M^4} {1\over2} Tr[ F \wedge * F ]
\end{equation}
that is just the usual Yang-Mills action on the curved spacetime
$M$.

It should be mentioned that we obtain the Lagrangian of Yang-Mills
theory simply by adding the {\it solutions} of constraints to the
original actions of the GTFT. This is only for the sake of
convenience. In first principle it is absolutely possible to add
the constraints themselves to the actions and then solve these
constraints.

For the Yang-Mills theory, % as the example,
we may add the constraints to the action:
\begin{eqnarray}
{\cal S}_{\rm{YM}}[{\cal A}]&=&\int_{M^4} {\cal L}_{\rm{YM}}
=\int_{M^4} \, {\bf F}^{p}\wedge {\bf F}_{p} +\int_{M^4}{\cal L}_{\rm{Constr.}}\nonumber\\
&=& \int_{M^4} \, F^{p} \wedge F_{p} + 2k F^{p}\wedge B_{p}+k^2
B^{p} \wedge B_{p}-2k^2 \phi^p_{AB}B_{p}\wedge B^{AB} .
\label{YM-action2}
\end{eqnarray}
where $\phi^p_{AB}$ is the Lagrangian multiplier. Varying the
action with respect to $B_{p}$ yields the following equations of
motion,
\begin{equation}
kB^{p}=k\phi^p_{AB}B^{AB}-F
\end{equation}
Substituting it into the action we then obtain the action
(\ref{CYM}) up to a topological term and a sign difference ahead
of the term $ \phi_{AB} \phi_{CD} B^{AB}\wedge B^{CD}$, which does
not affect any dynamics of Yang-Mills theory.
\section{N=1 Chiral Supergravity as Constrained Topological Field Theory}

In next two sections we apply the general formalism for $GTFT$ via
$GDC$ to construct the first-order formalism of $N=1, 2$ chiral
supergravities.

 As is know, the pure connection formulation of such kind of
supergravity theories originally were presented by virtue of
Ashtekar-Sen's variables in \cite{J} and \cite{GSU1},
respectively. The canonical quantization of $N=1,2$ supergravities
based on this framework were considered in \cite{GSU1,LS3} as
well. Furthermore, Ezawa argued in \cite{super-ezawa} that $N=1,2$
supergravities can be cast into the form of topological field
theories as for the case of GR. Here we will show that our general
framework for $GTFT$ via $GDC$ in the supersymmetric case provides
a more direct manner to understand the topological origin of this
formulation. In addition, our formalism also provides a convenient
and compact way to construct such kind of supergravities. In fact,
we find that by means of this formalism a class of topological
field theories with free parameters may be set up while the usual
known supergravities are the ones with the parameters specially
chosen.

In this section we derive the Lagrangian of $N=1$ chiral
supergravity with the $Osp(1|2)$ algebra, which is the simplest
supersymmetric extension of $su(2)$ algebra.
\begin{eqnarray}
\left[ J_{AB}, J_{CD} \right] &=&
      \epsilon_{C(A} M_{B)D}
     +\epsilon_{D(A} M_{B)C} \nonumber\\
\left[ J_{AB}, Q_C \right] &=&
      \epsilon_{C(A}Q_{B)}\nonumber\\
\left\{ Q_A, Q_B \right\} &=& a J_{AB},
\end{eqnarray}
where $a$ is a dimensionless constant to be fixed in order to give
a reasonable gravity theory. Later we will find it is related to
the cosmological constant as $a\sim -l^2\sqrt{-\Lambda}$. The
super connection 1-form valued on this algebra is defined as
\begin{equation}
{\cal A}=A^{AB}J_{AB}+\psi^AQ_A.
\end{equation}
In order to apply the $GDC$, we introduce a gauge covariant 2-form
field
\begin{equation}
{\cal B}=B^{AB}J_{AB}+B^AQ_A.
\end{equation}
Thus, we may define a generalized 1-form as
\begin{equation}
{\cal \bf A}=({\cal A}, {\cal B})=(A^{AB}, B^{AB})J_{AB}+(\psi^A,
B^A)Q_A.
\end{equation}
Following the connection theory via $GDC$, the generalized
curvature takes the form as
\begin{eqnarray}
{\cal \bf F} &=& {\bf d}{\cal\bf A}+{\cal \bf A}\wedge {\cal\bf A}\nonumber\\
         &=& (F^{AB}+kB^{AB}-{a\over 2}\psi^A\wedge \psi^B, DB^{AB}-a\psi^A\wedge B^B)J_{AB}\nonumber\\
         &+& (F^A+kB^A, DB^A)Q_A,
\end{eqnarray}
where
\begin{equation}
F^{AB}=dA^{AB}+A^{AC}\wedge {A_C}^{B}\;\;\;\;\;\;\;
F^A=d\psi^A+A^{AB}\wedge \psi_B.\end{equation} For the action, we
introduce the $GTFT$-type one (\ref{SFF-1})
\begin{eqnarray}\nonumber
S=\int_{ M}STr({\cal\bf F}\wedge{\cal\bf
F}).\nonumber\end{eqnarray} The 4-form components of its
Lagrangian read as
\begin{eqnarray}
STr({\cal \bf F}\wedge{\cal\bf F})&=&
 F^{AB}\wedge F_{AB}+2k F^{AB}\wedge B_{AB}+ k^2B^{AB}\wedge B_{AB}\nonumber\\&+&{a^2\over 4} \psi^A
                          \wedge\psi^B\wedge\psi_A\wedge\psi_B\nonumber\\
                          &-&
                          aF^{AB}\wedge\psi_{A}\wedge\psi_{B}-akB^{AB}\wedge\psi_{A}\wedge\psi_{B}\nonumber\\
                          &+& aF^A\wedge F_A + 2ak F^A\wedge B_A +
                          ak^2 B^A\wedge B_A,\label{SBF}
\end{eqnarray}
where we have used bilinear relations
\begin{equation}
STr{Q^AQ^B}=a\epsilon^{AB}, \;\;\;\;\;\;
STr{J_{AB}J^{CD}}=\delta_A^C\delta_B^D.\end{equation}

Now we return to the action (\ref{SBF}). This action can be
further simplified by reading off the ordinary topological terms.
Combining all the terms without $B$ fields in the action we easily
find that it is nothing but the ordinary supersymmetric version of
the second Chern class $STr{\cal F}\wedge {\cal F}$ with
supersymmetric curvature ${\cal F}$, which gives a total
derivative term in the sense of ordinary differential calculus,
i.e. $d{\cal L}_{CS}$\footnote{Of course among these terms we may
immediately delete the term with four $\psi$s due to the fact
$\psi^A\wedge \psi_A=0$, but it would be better to keep in mind
that such terms with four $\psi$s do not necessarily vanish in
$N>1$ supergravities since spinors are specified by more indices.
One example appearing in next section is that $\psi_i^A\wedge
\psi_{jA}=0$ only if $i,j$ is symmetric.}. If we read off all
these ordinary topological terms without $B$ fields, the remaining
Lagrangian becomes
\begin{eqnarray}
{\cal L} &=& 2k F^{AB}\wedge B_{AB} + k^2 B^{AB}\wedge
B_{AB}-akB^{AB}\wedge\psi_A\wedge\psi_B\nonumber\\
         &+&2ak F^A\wedge B_A+ ak^2 B^A\wedge B_A .
\end{eqnarray}
Now it is straightforward to identify this action with a kind of
the supersymmetric $BF$ theories with the freedom of rescaling the
coupling  constants $a,k$ and $B$ fields. Therefore, our action
(\ref{SBF}) of the $GTFT$ type via $GDC$ gives rise to a kind of
supersymmetric $BF$ theories up to an ordinary supersymmetric
version of the second Chern class that is a totally derivative
term. However, it should be noted that in our formalism both $a$
and $k$ are free parameters and not necessarily related to each
other. The $BF$ action given in \cite{super-ezawa} is obtained in
the special case of $k={-a^2\over 3}$ and $B'_A\sim aB_A$. In this
sense we generalize the original formalism into a class of
theories which of course should be topologically equivalent.

Let us now introduce the local degrees of freedom for both gravity
and gravitino fields by constraining the $B$ fields. This can be
done by either adding new terms $ \lambda_{(ABCD)}B^{AB}\wedge
B^{CD}$ and $\lambda_{(ABC)}B^{AB}\wedge B^C$ into the action or
simply plugging following solutions into the action,
\begin{eqnarray}
B_{AB}:&=& {e_A}^{A'}\wedge e_{A'B}\\
B_A:&=& {1\over a} {e_A}^{A'}\wedge\chi_{A'},
\end{eqnarray}
where $\lambda_{(ABCD)}$ and $\lambda_{(ABC)}$ are Lagrangian
multipliers with symmetrized indices.  After that we find a kind
of the Lagrangians with some free parameters as follows
\begin{eqnarray}
{\cal L} &=& 2k (F^{AB}\wedge {e_A}^{A'}\wedge e_{A'B} + {k\over
2}e^{AB'}\wedge {e^B}_{B'}\wedge {e_A}^{A'}\wedge e_{A'B}\nonumber\\ &-& {a\over 2}e^{AB'}\wedge {e^B}_{B'}\wedge\psi_A\wedge\psi_B\nonumber\\
         &+& F^A\wedge {e_A}^{A'}\wedge\chi_{A'}+ {k\over 2a} e^{AB'}\wedge \chi_{B'}\wedge
         {e_A}^{A'}\wedge\chi_{A'}).
\end{eqnarray}
Note that all these Lagrangians with arbitrary $k$ and $a$ are
$N=1$ chiral supersymmetric.

In order to give the CDJ formalism of $N=1$ chiral supergravity
\cite{CDJ}, we need fix the constants $k$ and $a$ as
\begin{equation}\label{ka}
k={\Lambda \over 3},\;\;\;\;\; \; a=-\sqrt{-\Lambda}
~.\end{equation} As a result, the final form of the known $N=1$
supergravity Lagrangian becomes
\begin{eqnarray}
{\cal L}^{sugra} &:=& {{\cal L}\over 2k}= (F^{AB}\wedge
{e_A}^{A'}\wedge e_{A'B} + {\Lambda \over 6}
e^{AB'}\wedge {e^B}_{B'}\wedge {e_A}^{A'}\wedge e_{A'B}\nonumber\\&+& {\sqrt{-\Lambda} \over 2}e^{AB'}\wedge {e^B}_{B'}\wedge\psi_A\wedge\psi_B\nonumber\\
         &+& F^A\wedge {e_A}^{A'}\wedge\chi_{A'}+ {\sqrt{-\Lambda}\over 6} e^{AB'}\wedge \chi_{B'}\wedge {e_A}^{A'}\wedge\chi_{A'})
\end{eqnarray}
We note here $\Lambda$ has to be negative due to the appearance of
square root. It simply means we only have a Lagrangian with
negative cosmological constant.

%June 24, added by Yi
At the remainder of this section we briefly discuss the theory in
the presence of boundary. It is well known that in this case we
need some boundary terms  so as to make the variation of the total
action well defined. One option is adding such boundary terms by
hand.

However, in our formalism one may have an alternative way to treat
the boundary terms. Rather than adding any boundary action by hand
as done in \cite{Smolinholo,LS2,Iso}, we may induce an ordinary
super Chern-Simons action on the boundary from the topological
term ${1\over 2k} STr {\cal F}\wedge {\cal F}$ in the bulk.  As we
noticed, it is nothing but the topological term
\begin{equation} STr({\cal F}\wedge {\cal F})=d {\cal
L}_{CS},\end{equation} where
\begin{eqnarray} {\cal L}_{CS}&=& STr ({\cal A} \wedge d{\cal A} +
{\textstyle{2\over3}} {\cal A} \wedge {\cal A} \wedge {\cal
A})\nonumber\\&&=A^{AB}\wedge dA_{AB}+{2\over 3}{A^A}_C\wedge
{A^C}_B\wedge {A^B}_A-a\psi^{ A}\wedge D\psi_{ A}.\end{eqnarray}
Thus in the presence of boundary, the total action of supergravity
is
\begin{equation}
S\equiv {S\over 2k}=\int_M {\cal L}^{sugra}+{1\over
2k}\int_{\partial M}(A^{AB}\wedge dA_{AB}+{2\over 3}{A^A}_C\wedge
{A^C}_B\wedge {A^B}_A-a\psi^A\wedge D\psi_A).\label{cscc}
\end{equation}Interestingly enough, from (\ref{cscc}) we find the coefficient of
the second Chern class ${\kappa\over 8\pi}$ should be fixed by the
cosmological constant as
\begin{equation} {\kappa\over
8\pi}={1\over 2k}={3\over 2\Lambda}.\end{equation} Furthermore, to
make the variation of the total action  well defined, we still
need impose the boundary condition, which should be
\begin{equation}
STr \delta {\cal A}\wedge {\cal B}={3\over \Lambda}STr \delta
{\cal A}\wedge {\cal F}.
\end{equation}

\section{N=2 Chiral Supergravity as Constrained Topological Field Theory}
In this section we demonstrate that our formalism for the $GTFT$
via $GDC$ can be extended to $N=2$ superalgebra in almost the same
manner as in previous sections and thus applicable to the
construction of a sort of $N=2$ supergravities with some free
parameters. The known $N=2$ supergravity is the one with the
parameters specially chosen.

The Lie super-algebra corresponding to $N=2$ chiral supergravity
is $Osp(2|2)$. It is composed of bosonic generators $J_{AB}$,$Q$
which span the algebra $sp(2)\times o(2)$, and fermionic
generators $Q_A^i(i=1,2)$. The algebraic relations read as
\begin{eqnarray}
\left[ J_{AB}, J_{CD} \right] &=&
      \epsilon_{C(A} M_{B)D}
     +\epsilon_{D(A} M_{B)C} \nonumber\\
\left[ J_{AB}, Q^i_C \right] &=&
      \epsilon_{C(A}Q^i_{B)}\nonumber\\
\left\{ Q^i_A, Q^j_B \right\} &=& a
J_{AB}\delta^{ij}+\epsilon_{AB}\epsilon^{ij}Q\nonumber\\
\left[ Q, Q_A^i\right] &=& {1\over 2}a\epsilon^{ij}Q_{Aj}\nonumber\\
\left[Q, Q\right] &=& \left[Q, J_{AB}\right]=0,
\end{eqnarray}
where we introduce the matrix $\epsilon^{ij}$ %and the third
%component of the Pauli matrices
\begin{equation}
\epsilon^{ij}=\left(
\begin{array}{cc}
  0 & 1 \\
  -1 & 0 \\
\end{array}
\right). \end{equation} To check the closure of this superalgebra
it is useful to
note the following identity %relations between these two matrices:
\begin{equation}
\epsilon^{ij}\epsilon^{km}=\delta^{jk}\delta^{im}-\delta^{ik}\delta^{jm}.
\end{equation}

The super connection 1-form now is defined as
\begin{equation}
{\cal A}=A^{AB}J_{AB}+\psi_{i}^AQ^i_A+A Q,
\end{equation}
where $A$ is a $U(1)$ connection. In parallel we introduce a gauge
covariant 2-form field
\begin{equation}
{\cal B}=B^{AB}J_{AB}+B_{i}^AQ^i_A+BQ.
\end{equation}

The generalized connection 1-form is defined as the pairing
\begin{equation}
{\cal \bf A}=({\cal A}, {\cal B})=(A^{AB},
B^{AB})J_{AB}+(\psi_i^A, B_i^A)Q^i_A+(A,B)Q.
\end{equation}
Thus the generalized curvature 2-form are given by %takes the form as
\begin{eqnarray}
{\cal \bf F} &=& {\bf d}{\cal\bf A}+{\cal \bf A}\wedge {\cal\bf A}\nonumber\\
         &=& (F^{AB}+kB^{AB}-{a\over 2}\psi^{iA}\wedge \psi_i^B,
         DB^{AB}-a\psi^{iA}\wedge B_i^B)J_{AB}\nonumber\\
         &+& (F_i^A+kB_i^A+{1\over2}a\epsilon_{ij}A\wedge \psi^{jA},
         DB_i^A+a\epsilon_{ij}A\wedge B^{jA})Q^i_A\nonumber\\
         &+& (dA+kB+{1\over 2} \epsilon^{ij}\psi_i^A\wedge \psi_{jA}, dB+\epsilon^{ij}\psi_i^A\wedge
         B_{jA})Q.
\end{eqnarray}
It satisfies the Bianchi identity with respect to $\Gd$.

The generalized topological action can be constructed with the use
of this supercurvature 2-form as
\begin{equation}
S=\int_{ M}STr{\cal\bf F}\wedge{\cal\bf F}\end{equation} where
\begin{equation}
STr{Q^i_AQ^j_B}=a\epsilon_{AB}\delta^{ij}, \;\;\;\;
STr{J_{AB}J^{CD}}=\delta_A^C\delta_B^D, \;\;\;\;
STr{QQ}=a^2.\end{equation} More explicitly its 4-form components
read as
\begin{eqnarray}
STr{\cal \bf F}\wedge{\cal\bf F}&=&
 F^{AB}\wedge F_{AB}+2k F^{AB}\wedge B_{AB}+ k^2B^{AB}\wedge B_{AB}\nonumber\\&+&{a^2\over 4}
 \psi^{iA}\wedge\psi_i^B\wedge \psi^j_{A}\wedge\psi_{jB}\nonumber\\
                          &-&aF^{AB}\wedge \psi^i_{A}\wedge\psi_{iB}
                          -akB^{AB}\wedge \psi^i_{A}\wedge\psi_{iB}\nonumber\\
                          &+& a[ F^{iA}\wedge F_{iA} + 2k F^{iA}\wedge B_{iA} +
                          k^2 B^{iA}\wedge B_{iA} \nonumber\\
                           &+& akB^{iA}\wedge A \wedge \epsilon_{ij}\psi^j_{A}+{a^2\over 4}A\wedge \psi^{iA}\wedge A
                           \wedge \psi_{iA}
                          +aF^{iA}\wedge A\wedge \epsilon_{ij}\psi^j_{A}]\nonumber \\
                          &+& a^2[ F\wedge F +k^2B\wedge B +2k F\wedge B+ k \epsilon^{ij}\psi_i^A\wedge\psi_{jA}\wedge B\nonumber\\
   &+& F\wedge \epsilon^{ij}\psi_i^A\wedge \psi_{jA}+{1\over 4}\epsilon^{ij}\psi_i^A\wedge \psi_{jA}
   \wedge \epsilon^{mn}\psi_m^B\wedge \psi_{nB}],\label{SBF2}
\end{eqnarray}
Again, we separate all the terms without $B$ fields and combine
them together, leading to an ordinary topological term $STr{\cal
F}\wedge {\cal F}$ where ${\cal F}$ is the $Osp(2|2)$-valued
curvature 2-form. Reading off this topological term, we obtain the
following $BF$ type Lagrangians with free parameters $k$, $a$ as
follows:
\begin{eqnarray}
{\cal L} &=& 2k (F^{AB}\wedge B_{AB} + {k\over 2} B^{AB}\wedge
B_{AB}-{a\over 2}B^{AB}\wedge \psi^i_{A}\wedge\psi_{iB}\nonumber\\
         &+&a F^{iA}\wedge B_{iA}+ {ak\over 2} B^{iA}\wedge B_{iA}
         +{a^2\over 2}B^{iA}\wedge A \wedge \epsilon_{ij}\psi^j_A\nonumber \\
         &+& {ka^2\over 2 }B\wedge B+a^2 dA\wedge B+ {a^2\over 2}B \wedge \epsilon^{ij}\psi_i^A\wedge
         \psi_{jA}).
\end{eqnarray}
We are free to rescale the $B$ fields as $B_i^A\sim {1\over
a}B_i^A$, $B\sim {1\over a^2}B$. Furthermore, to identify our
Lagrangian with that in \cite{super-ezawa}, we redefine the
coupling constants %
\begin{eqnarray}\label{N2ak} a=2g,% and require
\quad k=2g^2={-\Lambda \over 3 }.
\end{eqnarray}
 As a result,
\begin{eqnarray}
{{\cal L}\over 2k} &=& F^{AB}\wedge B_{AB} + g^2 B^{AB}\wedge
B_{AB}-gB^{AB}\wedge \psi^i_{A}\wedge\psi_{iB}\nonumber\\
         &+& F^{iA}\wedge B_{iA}+ {g\over 2} B^{iA}\wedge B_{iA}
         +gB^{iA}\wedge A \wedge \epsilon_{ij}\psi^j_A\nonumber \\
         &+& {1\over 4} B\wedge B+ \hat{F}\wedge B.\label{SBF3}
\end{eqnarray}
where
\begin{equation}
\hat{F}=dA+{1\over 2}\epsilon^{ij}\psi_i^A\wedge
         \psi_{jA}
\end{equation}
The supergravity Lagrangian is obtained by plugging the following
constraints into (\ref{SBF3}):
\begin{equation}
B_{AB}={e_A}^{A'}\wedge e_{BA'}\;\;\;\;\; B_{Ai}={e_A}^{A'}\wedge
\chi_{iA'},
\end{equation}
and
\begin{equation}
{e_A}^{A'}\wedge e_{BA'}\wedge B={e_A}^{A'}\wedge \chi_{iA'}\wedge
{e_B}^{B'}\wedge \chi^i_{B'},
\end{equation}

The subtlety that the last constraint indeed gives the Maxwell
term of $U(1)$ field can be understood following our discussion in
Yang-Mills section. Namely after plugging the constraints into the
action of $BF$ theory, we obtain the terms involving the auxiliary
field $B$ in action (\ref{SBF3}) as
\begin{equation}
{\cal L}_{U(1)}={1\over 4} B\wedge B+ \hat{F}\wedge
B+\phi_{AB}(B^{AB}\wedge B-e_A^{A'}\wedge \chi_{iA'}\wedge
e_B^{B'}\wedge \chi^i_{B'}).
\end{equation}
Varying with $B$ yields the equation of motion
\begin{equation}
{1\over 2} B=- \hat{F}-\phi_{AB}B^{AB}.
\end{equation}
Substituting this solution back to the action immediately leads to
the chiral action of $N=2$ supergravity given in \cite{GSU1}.

Note that this known chiral action of $N=2$ supergravity is just
the special case with the chosen parameters (\ref{N2ak}). In fact,
as in the case of $N=1$ chiral supergravities, a kind of $N=2$
chiral supergravities with free parameters should also be set up
in general in our formalism of the $GTFT$ with constraints via
$GDC$.

\section{Concluding Remarks}

In this paper, we have further explored our approach  to the
connection theory and $GTFT$ with constraints via the $GDC$
\cite{BFT} to both the Einstein-Hilbert action in GR  and
Yang-Mills action on curved spacetime. We have also extended it as
a general formalism including the supersymmetric cases. In
particular we have constructed $N=1,2$ supersymmetric
$BF$-theories with free parameters as well as chiral
supergravities as the parameters are specially chosen. The common
properties of such kinds of $GTFTs$ are of the $BF$-type
companying with the ordinary second Chern-class in the bulk and
the Chern-Simons term on the boundary. This not only has recovered
the deep relation between the $BF$-type $TFT$ and the Chern-Simons
one, but also has combined them together in the bulk or on the
boundary, respectively, rather than added by hand as has been done
in literatures. In all these cases, Ashtekar-Sen's variables have
been employed. This is quite convenient, although it is not
necessary.

It should be noticed that such a general formalism works only for
the (super)gravity theories with non-zero cosmological constants.
This is much similar to the case that Chern-Simons states as the
topological solutions to quantum GR exist only in the case of
non-zero cosmological constants. This similarity is sharpen if we
observe the total action (\ref{cscc}) where the coefficients of
Chern-Simons action on the boundary has to be fixed inversely
proportional to the cosmological constants as well as Chern-Simons
states in quantum gravity. As we know the Chern-Simons theory on
the boundary plays the key role in the holographical formulation
of GR at quantum mechanical level \cite{Smolinholo, LS2}, it is
reasonable to conjecture that the holographical interpretation of
quantum gravity make sense only for theories with non-zero
cosmological constants.

It is worthwhile to remind  that the basic properties of four
dimensional $BF$ theory have previously been studied in many
references, for instance see \cite{BBRT,Horowitz,CCFM}.  Recently
it turns out that $BF$ formulation of GR plays a crucial role in
the quantization of gravitational fields as well. Some of its
important applications may be illustrated  as follows.

First one is to study the evolution of spacetime at quantum
mechanical level. Originally Crane and Yetter proposed that using
$BF$ theory the algebraic method in topological quantum field
theory can be applied to construct quantum geometry in the form of
a discrete model on a triangulated 4-manifold \cite{Crane,CY}.
This idea is quite similar to the Ponzano-Regge state sum model
for three dimensional gravity, but one significant difference here
is that gravity has the local degrees of freedom in four
dimensions, which requires imposing some additional constraints to
$BF$ Lagrangian. After the appearance of spin networks in loop
quantum gravity, these ideas ware quickly applied to the
construction of state sum model \cite{BC} and spin foams
\cite{RR,Baez}.

The second application is to construct holographic formulation of
non-perturbative quantum gravity and
supergravity\cite{Smolinholo,LS2}, motivated by the original work
of 't Hooft and Susskind, who roughly speaking conjecture that in
quantum gravity the state space describing physics in a region
with finite volume should be finite dimensional. The basic idea
here is to study the quantization of gravity in the presence of
finite boundary. It turns out that $BF$ formulation of gravity in
this context provides not only an ideal framework for imposing the
appropriate boundary conditions on spacetime, but also the key
mathematical structure to construct the boundary theory of quantum
gravity.

The third application is the study of isolated horizons\cite{Iso}.
Given a $BF$ formulation of gravity in the bulk, the strategy of
describing quantum gravity on the isolated horizons can be carried
out closely following the idea in \cite{Smolinholo}, while the
main different ingredient here is the treatment of boundary
conditions.

It is needless to say that in all these applications our general
formalism may play some essential role when we want to extend
above considerations to supergravity theories.

In our previous paper \cite{BFT} and this paper, the 4-dimensional
manifolds are concerned. In principle, our approach may apply to
other dimensions. Although it seems that the action of GR in
Ashtekar-Sen variables may not be applicable for the manifolds
rather than four dimensions, some kinds of $BF$ actions may appear
together with topological one as the candidates for the $GTFT$. On
the other hand, as was mentioned in the context, there may exist a
kind of decent relations among the topological terms on different
dimensions since in the action of $GTFT$ the topological terms of
different forms appear already via $GDC$.

Furthermore we notice that the concept of a generalized p-form
discussed in our paper is only a special case of a broader
generalization in which generalized p-forms may be constructed by
$(n+1)$-tuples of ordinary forms \cite{NR2001,Robinson}. It is of
particular interest to investigate the physical application to
higher dimensional supergravity theories where generalized p-form
connections are allowed to associate to $(p-1)$-branes.

\begin{acknowledgments}
The authors would like to thank Professor Jianzhong Pan for
valuable discussion on $GDC$. This work is partly supported by
NSFC (Nos. 90103004, 10175070). Y. Ling is partly supported by
NSFC (No. 10205002) and K.C.Wong Education Foundation, Hong Kong.
\end{acknowledgments}
\section*{Appendix: Generalized Differential Calculus}

A generalized $p$-form \cite{Sparling,NR2001}, $\GForm{p}{a}$, is
defined to be an ordered pair of an ordinary $p$-form
$\Form{p}{\alpha}$ and an ordinary $(p+1)$-form
$\Form{p+1}{\alpha}$ on an n-dimensional manifold $M$, that is
\begin{equation}
\GForm{p}{a} \equiv (\Form{p}{\alpha},\Form{p+1}{\alpha}) \in
\Lambda^p \times \Lambda^{p+1} ,
\end{equation}
where $-1 \leq p \leq n$. The minus one-form is defined to be an
ordered pair
\begin{equation}
\GForm{-1}{a}\equiv (0, \Form{0}{\alpha}) ,
\end{equation}
where $\Form{0}{\alpha}$ is a function on $M$. The product and
derivatives are defined by
\begin{equation}
\GForm{p}{a} \wedge \GForm{q}{b}\equiv ( \Form{p}{\alpha}\wedge
\Form{q}{\beta}, \Form{p}{\alpha} \wedge \Form{q+1}{\beta}+(-1)^q
\Form{p+1}{\alpha}\wedge \Form{q}{\beta}) ,
\end{equation}
\begin{equation}
\Gd \GForm{p}{a} \equiv ( d \Form{p}{\alpha}+(-1)^{p+1} k
\Form{p+1}{\alpha}, d \Form{p+1}{\alpha}) ,
\end{equation}
where $k$ is a constant. These exterior products and derivatives
of generalized forms satisfy the standard rules of exterior
algebra
\begin{equation}
\GForm{p}{a} \wedge \GForm{q}{b} = (-1)^{pq} \GForm{q}{b} \wedge
\GForm{p}{a} ,
\end{equation}
\begin{equation}
\Gd (\GForm{p}{a} \wedge \GForm{q}{b}) = \Gd \GForm{p}{a} \wedge
\GForm{q}{b} + (-1)^{p} \GForm{p}{a} \wedge \Gd \GForm{q}{b} ,
\end{equation}
and ${\bf d}^2=0$.

For a generalized p-form $\GForm{p}{a}(\Form{p}{\alpha},\Form{p+1}{\alpha})$, the integration on $M^p$
can be defined as usual by
\begin{equation}
\int_{M^p} \GForm{p}{a}=\int_{M^p}
(\Form{p}{\alpha},\Form{p+1}{\alpha}) = \int_{M^p}
\Form{p}{\alpha} .
\end{equation}

The generalized connection and curvature on $P(M, G)$ have been
introduced and the generalized Chern-Weil homomorphism for
generalized curvature invariant polynomials in any even
dimensional manifolds have also been established. But, their
topological meaning should be as same as before. This can be
understood by the fact that for $\Gd$, the cohomology is trivial.
We show this  property as follows. Given a generalized $p$-form
which is closed in the context of GDC, namely
%\begin{eqnarray}*
 \begin{equation}{\bf d}(\Form{p}{\alpha},\Form{p+1}{\beta})
 =(d\Form{p}{\alpha}+(-1)^{p+1}k\Form{p+1}{\beta},~d\Form{p+1}{\beta})=0,\end{equation}
%\end{eqnarray}*
  then we have \begin{equation} d\Form{p}{\alpha}+(-1)^{p+1}k\Form{p+1}{\beta}=0,\quad
  d\Form{p+1}{\beta}=0\end{equation} for any non-zero constant
  $k$.
 %\begin{eqnarray}*
 Thus,
 \begin{equation}\Form{p+1}{\beta}=(-1)^{p}k^{-1}d\Form{p}{\alpha}.\end{equation}
 It follows
$$(\Form{p}{\alpha},~(-1)^{p}k^{-1}d\Form{p}{\alpha})=\Gd
(0,~(-1)^{p}k^{-1}\Form{p}{\alpha}).
 $$%\end{eqnarray}*
Namely, the closed generalized $p$-form is always exact. Therefore
the cohomology is trivial.

This property implies that the ordinary $BF$ theory combined with
the ordinary  second Chern class may share the same topological
meaning with the ordinary second Chern class. The generalized
topological field theory just provides a framework for the
combination of these two sorts of theories. Then it is quite
natural to see that for an ordinary $BF$ theory in the bulk, we
may have a Chern-Simons action induced on the boundary in this
context.

\end{document}